\documentclass[letterpaper,12pt]{article}
\usepackage{color,amssymb,amsmath,enumerate,float}

\usepackage{ifpdf}
\ifpdf
	\usepackage[pdftex]{graphicx}
	\usepackage[pdftex,unicode,implicit]{hyperref}

	\hypersetup{
  	pdftitle     = {Einstein's quadrupole}, 
  	pdfkeywords  = {},
  	pdfauthor    = {},
  	pdfcreator   = {pdf\LaTeXe\ with package \flqq hyperref\frqq},
  	pdfproducer  = {pdf\LaTeXe\ with package \flqq hyperref\frqq},
  	pdfpagemode  = UseNone,  
  	pdffitwindow = true,  
  	unicode      = true,
  	plainpages   = true,
  	colorlinks   = true,  
  	citecolor    = black,  
  	urlcolor     = blue, 
  	linkcolor    = black
	}

\else

  \usepackage[dvips]{graphicx}

\fi 

\makeatletter
\@addtoreset{equation}{section}
\makeatother

\begin{document}
	
\thispagestyle{empty}

\begin{center}

{\bf \LARGE Einstein's quadrupole formula from the kinetic-conformal Ho\v{r}ava theory}
\vspace*{15mm}

{\large Jorge Bellor\'{\i}n}$^{a,1}$
{\large and Alvaro Restuccia}$^{a,b,2}$
\vspace{3ex}

{\it $^a$Department of Physics, Universidad de Antofagasta, 1240000 Antofagasta, Chile.}
{\it $^b$Department of Physics, Universidad Sim\'on Bol\'{\i}var, 1080-A Caracas, Venezuela.} 
\vspace{3ex}

$^1${\tt jbellori@gmail.com,} \hspace{1em}
$^2${\tt arestu@usb.ve}

\vspace*{15mm}
{\bf Abstract}
\begin{quotation}{\small
   We analyze the radiative and nonradiative linearized variables in a gravity theory within the familiy of the nonprojectable Ho\v{r}ava theories, the Ho\v{r}ava theory at the kinetic-conformal point. There is no extra mode in this formulation, the theory shares the same number of degrees of freedom with general relativity. The large-distance effective action, which is the one we consider, can be given in a generally-covariant form under asymptotically flat boundary conditions, the Einstein-aether theory under the condition of hypersurface orthogonality on the aether vector. In the linearized theory we find that only the transverse-traceless tensorial modes obey a sourced wave equation, as in general relativity. The rest of variables are nonradiative. The result is gauge-independent at the level of the linearized theory. For the case of a weak source, we find that the leading mode in the far zone is exactly Einstein's quadrupole formula of general relativity, if some coupling constants are properly identified. There are no monopoles nor dipoles in this formulation, in distinction to the nonprojectable Horava theory outside the kinetic-conformal point. We also discuss some constraints on the theory arising from the observational bounds on Lorentz-violating theories.
}
\end{quotation}

\end{center}

\thispagestyle{empty}

\newpage
\section{Introduction}
Gravitational waves have recently been detected \cite{Abbott:2016blz,Abbott:2016nmj,Abbott:2017vtc}. The detected signals fit well with the waves produced by the coalescense of binary systems of black holes, according to the predictions of General Relativity (GR). This detection constitutes one of the most important recent achievements in the study of gravitational phenomena, and it is another success of GR. On the other hand, there are motivations to study alternatives or modifications to GR. One important issue is that GR is not renormalizable under the perturbative approach, thus, at least in the perturbative scheme, it cannot be a fundamental theory by itself. There is also the issue of the dark matter, for which there has not been found any candidate in the particle experiments or space observations. Therefore, for any proposed modification of GR a question comes out inmediatly: how close to the detected wave signals and the corresponding predictions of GR is the radiation predicted by the new theory?.

Here we focus on the study of the production and propagation of gravitational waves at the leading order in a Lorentz-violating theory. The theory \cite{Bellorin:2013zbp} belongs to the family of the nonprojectable Ho\v{r}ava theories \cite{Horava:2009uw,Blas:2009qj}. The heart of the Ho\v{r}ava proposal \cite{Horava:2009uw} is to introduce a preferred timelike direction that breaks the symmetry between space and time characteristic of relativistic theories. This is done with the aim of introducing higher order spatial derivatives in the Lagrangian that improve the renormalizability of the theory while, in principle, preserve its unitarity. The special formulation studied in Ref.~\cite{Bellorin:2013zbp}, where only the purely gravitational theory without coupling to matter sources was analyzed, consists of setting a specific value of the kinetic coupling constant for which two additional second-class constraints emerge. The constant is usually denoted by $\lambda$ and the special value in $3+1$ dimensions is $\lambda = 1/3$. The additional constraints at $\lambda = 1/3$ eliminate the extra scalar mode that otherwise the nonprojectable Ho\v{r}ava theory exhibits. Because of this, it is reasonable to expect that this formulation tends to stay more close to GR, at least in the low-energy regime, where the lowest order operators are the most relevant ones\footnote{The $U(1)$ gauge extensions, both in the projectable \cite{Horava:2010zj} and nonprojectable \cite{Zhu:2011xe} versions of the Ho\v{r}ava theory, also eliminate the extra mode.}. Since the special value $\lambda =1/3$ is related to a conformal symmetry on the kinetic term of the Lagrangian \cite{Horava:2009uw}, in Ref.~\cite{Bellorin:2016wsl} we called this formulation the Ho\v{r}ava theory at the kinetic-conformal point (the KCP Ho\v{r}ava theory, for short). We stress that the theory is not conformally invariant, only its kinetic term is. In \cite{Bellorin:2016wsl} the power-counting renormalizability as well as the absence of ghosts in the theory were shown. A recent report on the status of the Ho\v{r}ava theory, dealing with its several versions, can be found in Ref.~\cite{Wang:2017brl}. We comment that the value $\lambda = 1/3$ leading to the kinetic-conformal formulation is fixed by the dynamics, it does not get quantum corrections. This is due to the second-class constraints of the theory, not to symmetries. Further discussion can be found in Ref.~\cite{Bellorin:2017gzj}.

We study the  gravitational waves at leading order in the large-distance effective action of the theory. This effective action is of second order in time and spatial derivatives. It admits a generally-covariant version which is the Einstein-aether theory \cite{Jacobson:2000xp} under the restriction of hypersurface orthogonality on the aether vector \cite{Blas:2009ck,Jacobson:2010mx}. We deal with the generally-covariant formulation since it allows a more direct comparison with the standard approaches of GR devoted to gravitational waves. 

We analyse the perturbatively linearized generally-covariant theory coupled to a generic weak matter source. We develop all the analysis in terms of gauge-invariant variables of the linearized theory. These are combinations of the metric and the aether field that remain invariant under linearized general diffeomorphisms. This formulation allows us to get totally gauge-invariant results. Once we determine what variables are related to the sources by Poisson equations, hence they are nonradiative, and what variables are governed by the wave equation, we study the generation of the waves at the leading order. We follow the standard procedure of approximating the solution by considering that it is produced by a source that at the leading order has negligible self-gravity, it is in a slow motion regime, and that the observation is made far enough from the source (at the wave zone).

Our study is close to Refs.~\cite{Jacobson:2004ts,Foster:2006az,Blas:2011zd}. In Ref.~\cite{Jacobson:2004ts} a perturbative analysis of the unrestricted Einstein-aether theory without matter sources was done. The linearized vacuum equations of motion of the several modes, which are homogeneous wave equations for each mode with different speeds, were studied there. In Ref.~\cite{Foster:2006az} the lowest order multipole moments were studied for the unrestricted Eintein-aether theory coupled to a weak source. Our analysis differs from these two studies due to (besides the absence of sources in \cite{Jacobson:2004ts}) the lower number of propagating degrees of freedom we have and the fact that the equations of motion of the nonprojectable Ho\v{r}ava theory are not equivalent to the ones obtained by substituting the hypersurface orthogonality condition in the equations of motion of the Einstein-aether theory \cite{Jacobson:2010mx}.

In Ref.~\cite{Blas:2011zd} the Einstein-aether theory with the condition of hypersurface orthogonality imposed at the level of the action was studied (this theory is also called the khronometric theory). The analysis is rigorously consistent for $\lambda \neq 1/3$. In \cite{Blas:2011zd} the wave equations with sources were found for the tensorial modes and the extra mode, as well as the Poissonian equations for the nonradiative modes. They also found the dominant modes in the multipolar expansion for a weak source. These results are affected by the presence of the extra mode. Our study differs from the one of Ref.~\cite{Blas:2011zd} because we take the KCP theory independently with its intrinsic degrees of freedom. As we shall see, this has important consequences on the radiation formulas. In general, the KCP formulation cannot be obtained rigurously as a limit of the theory with the extra mode due to the discontinuity in the number of constraints and, in particular, in the number of propagating modes. Our approach is consistent in the $\lambda =1/3$ case since we obtain the formulas of the radiation directly from the KCP theory. However, if one wants a quick comparison with the non-kinetic-conformal case, heuristically our formulas coincide with the radiation formulas of Ref.~\cite{Blas:2011zd} in the limiting case of sending to infinity the speed of the extra mode (this divergence is induced by the $\lambda =1/3$ value). However, in general the reinterpretation of an hyperbolic equation (the wave equation of the extra mode) as an elliptic equation is not consistent (for example, the initial data problem).

In addition to the study of the gravitational radiation, here we consider some observational implications on the kinetic-conformal theory. Our aim is to highlight that some of the observational bounds applicable to the Ho\v{r}ava theory with $\lambda \neq 1/3$ must be addressed in a different way in the kinetic-conformal case. In Ho\v{r}ava theory observational bounds are frequently combined with theoretical restrictions needed for the consistency of the extra mode. In the kinetic-conformal theory this is not necessary since there is no extra mode. Another important issue is the cosmological-scale solutions. We have argued that they may arise in a different way in the kinetic-conformal case \cite{Bellorin:2017gzj}. We further comment on this point below. In particular, here we compare with the observational bounds coming from binary pulsars found in Ref.~\cite{Yagi:2013qpa,Yagi:2013ava}, since these phenomena are related to wave production. Those authors studied the Einstein-aether theory both unrestricted and with the hypersurface orthogonality condition (with $\lambda \neq 1/3$). They found stringent constraints on the space of coupling constants of these theories. Here we show how the kinetic-conformal theory stays more close to GR, in particular there is no dipolar contribution at the level of the dominant modes.

This paper is organized as follows: in section 2 we summarize the analysis of Ref.~\cite{Jacobson:2010mx} to present the Einstein-aether theory under the restriction of hypersurface orthogonality, together with its equivalence to the second-order action of the nonprojectable Ho\v{r}ava theory. In section 3.1 we discuss the gauge invariants of the generally-covariant theory that can be formed by combining the metric variables with the hypersurface-orthogonal aether field. In section 3.2 we present and analyze the linearized field equations, coupled to a matter source, in terms of these gauge invariants. In section 3.3 we study the leading mode for the production of waves far from the source, obtaining the quadrupole formula of Einstein. In section 4 we discuss some observational bounds. Finally we present some conclusions. Since it is also interesting to analyze the linearized field equations in the FDiff-covariant language, which is the original formulation of the Ho\v{r}ava theory \cite{Horava:2009uw}, we add one appendix to present the FDiff-gauge invariants and the field equations in terms of them. 


\section{The covariant version of the Ho\v{r}ava theory}
The Einstein-aether theory \cite{Jacobson:2000xp} is a modification of GR that incorporates an everywhere timelike unit vector field, called the aether, as a fundamental field. Since the aether is considered dynamical, the action possesses the symmetry of general diffeomorphisms that is also present in GR. However, at the level of the solutions, the presence of the aether field breaks the local Lorentz symmetry. There is a relationship \cite{Blas:2009ck,Jacobson:2010mx} between the Einstein-aether theory and the action of second order in derivatives of the nonprojectable Ho\v{r}ava theory, which is also a theory with a preferred frame.  Throghout this paper we deal only with the second-order action (excluding the cosmological constant), since at low precision the physics of the gravitational waves can be described by it. In the following we summarize the relationship between these two theories.

The Ho\v{r}ava theory \cite{Horava:2009uw} was originally formulated in terms of the standard Arnowitt-Deser-Misner (ADM) variables $N$, $N_i$ and $g_{ij}$, in such a way that the action possesses the symmetry of the diffeomorphisms that preserve a given foliation (FDiff) along a timelike direction. The Lagrangian of the nonprojectable theory, which is the version we study here, depends on the spatial curvature and the spatial derivatives of the lapse function $N$. These arise in the Lagrangian in terms of the FDiff-covariant vector $a_i = \partial_i \ln N$ \cite{Blas:2009qj}. The action of second order in derivatives, which we call the $z=1$ action, is
\begin{equation}
S = 
\frac{1}{2\kappa_{\mbox{\tiny H}}} \int dt d^3x \sqrt{^{(3)}\!g} N \left(
K_{ij} K^{ij} - \lambda K^2 + \beta\,^{(3)}\!R + \alpha a_i a^i \right) \,,
\label{horavaactiongeneral}
\end{equation}
where 
\begin{equation}
K_{ij} \equiv \frac{1}{2N} \left( 
\dot{g}_{ij} - 2 \nabla_{(i} N_{j)} \right)
\end{equation}
is the extrinsic curvature of the spacelike leaves $\Sigma$ of the foliation. The dot denotes the time derivative, $\dot{g}_{ij} = \partial g_{ij} / \partial t$. $K \equiv g^{ij} K_{ij}$, $^{(3)}\!R$ is the scalar curvature of $\Sigma$, and $\kappa_{\mbox{\tiny H}}$, $\lambda$, $\beta$ and $\alpha$ are coupling constants. 

In the above all the coupling constants are in principle arbitrary. Now, in the purely gravitational theory it is known that a scalar degree of freedom, additional to the transverse-traceless tensorial modes, is eliminated from the phase space if the coupling constant $\lambda$ is set to the value $\lambda = 1/3$ \cite{Bellorin:2013zbp,Bellorin:2016wsl}. With this value of $\lambda$ the kinetic term in (\ref{horavaactiongeneral}) acquires a conformal invariance \cite{Horava:2009uw}, although the full theory is not conformal since in general the terms in the potential break the conformal symmetry (except for very specific terms). For this reason the value $\lambda = 1/3$ was called the kinetic-conformal point in Ref.~\cite{Bellorin:2016wsl}. As we have mentioned, this feature raises interest in studying this special formulation of the nonprojectable Ho\v{r}ava theory, as it is our case in this paper, since it becomes closer to GR.

The Einstein-aether theory \cite{Jacobson:2000xp} is physically equivalent to the $z=1$ nonprojectable Ho\v{r}ava theory (\ref{horavaactiongeneral}) (for all $\lambda$) if the aether vector is restricted to be hypersurface orthogonal. The total equivalence between the two theories, at the level of their Lagrangians, holds only if the restriction on the aether vector is imposed at the level of the action, i. e. before deriving the equations of motion \cite{Blas:2009ck,Jacobson:2010mx}. Here we take the Einstein-aether action from Ref.~\cite{Jacobson:2010mx}, considering also the coupling to matter sources. The full generally-covariant action is given by $S_{\mbox{\tiny Total}} = S_{\mbox{\tiny EA}} + S_{\mbox{\tiny Matter}}$, where
\begin{equation}
 S_{\mbox{\tiny EA}}[ g_{\mu\nu} , u_\mu ] = 
 \frac{1}{2 \kappa_{\mbox{\tiny EA}} }\int d^4x \sqrt{-g} \left( 
 R - M^{\alpha\beta\gamma\delta} \nabla_\alpha u_\gamma 
           \nabla_\beta u_\delta \right)
\label{actioneinsteinaether2}
\end{equation}
is the Einstein-aether action. $u_\mu$ is the aether vector, which in general is subject to the condition of being a timelike unit vector, $u_\mu u^\mu = -1$. $\kappa_{\mbox{\tiny EA}}$ is the Einstein-aether gravitational constant. $M^{\alpha\beta\gamma\delta}$ is the hypermatrix
\begin{equation}
 M^{\alpha\beta\gamma\delta} =
 c_1 g^{\alpha\beta} g^{\gamma\delta} 
 + c_2 g^{\alpha\gamma} g^{\beta\delta}
 + c_3 g^{\alpha\delta} g^{\beta\gamma} 
 + c_4 u^\alpha u^\beta g^{\gamma\delta} \,,
\end{equation}
where $c_1$, $c_2$, $c_3$ and $c_4$ are coupling constants. With the aim of minimizing Lorentz-breaking effects in the matter sector, where experimental bounds are highly restrictive, it is required that the matter sources do not couple to the aether field (see discussion in Refs.~\cite{Foster:2005dk,Foster:2006az}). Then, $S_{\mbox{\tiny Matter}}[g_{\mu\nu},\psi]$ is the action for the matter sector, with $\psi$ representing the matter sources in a generic way. As a consequence, the equations of motion of the sources maintain the same structures they have in GR. 

The restriction of hypersurface orthogonality on $u_\mu$ is equivalent (locally) to express $u_\mu$ in terms of a scalar function $T = T(t,\vec{x})$ that satisfies the condition of its gradient is timelike, $\partial_\alpha T \partial^\alpha T < 0$. The hypersurface-orthogonal aether vector is written in terms of $T$ as
\begin{equation}
 u_\mu = 
 \frac{ \partial_\mu T }{ \sqrt{ - \partial_\alpha T \partial^\alpha T }} \,.
\label{uorthogonal}
\end{equation}
Under this restriction the functional degrees of freedom originally contained in $u_\mu$ are reduced to the one of $T$ once (\ref{uorthogonal}) has been substituted in the action (\ref{actioneinsteinaether2}). Actually, definition (\ref{uorthogonal}), which automatically implies that $u_\mu$ is a timelike unit vector, depends on the norm of the gradient of $T$, hence the hypersurface-orthogonal $u_\mu$ is a composite object made with the $T$ field and the metric $g_{\alpha\beta}$.

The equation of motion of the $T$ field, that is, the equation of motion derived from (\ref{actioneinsteinaether2}) by taking variations with respect to $T$, is implied by the Einstein equations and the matter equations of motion \cite{Jacobson:2010mx}. Since this fact is crucial for our study, let us repeat the argument that supports it. The main point  is that $T$ is a single, nonzero-gradient, scalar field coupled to gravity in a generally-covariant way. Since $S_{\mbox{\tiny Total}}$ is invariant under general diffeomorphisms, we have that, under a diffeomorphism parameterized by $\zeta^\mu$,
\begin{equation}
 0 = \int d^4x \left( 
 \frac{\delta S_{\mbox{\tiny Total}}}{\delta g_{\mu\nu}} 
    \mathcal{L}_\zeta g_{\mu\nu} 
 + \frac{\delta S_{\mbox{\tiny Total}}}{\delta T} \mathcal{L}_\zeta T
 + \frac{ \delta S_{\mbox{\tiny Total}}}{\delta \psi}
    \mathcal{L}_\zeta \psi \right) \,.
\label{invarianceaction}
\end{equation}
Now suppose that this identity is evaluated on configurations that satisfy the Einstein and matter equations. Over such configurations identity (\ref{invarianceaction}) becomes
\begin{equation}
 0 = \int d^4x 
   \frac{\delta S_{\mbox{\tiny Total}}}{\delta T} \mathcal{L}_\zeta T \,.
\end{equation}
Since $T$ cannot be constant along all possible directions and the above condition must be satisfied by all vectors $\zeta^\mu$, whe have that $\delta S_{\mbox{\tiny Total}} / \delta T = \delta S_{\mbox{\tiny EA}} / \delta T = 0$ automatically for all configurations that satisfy the Einstein equations and the matter equations of motion.

The physical equivalence between the action (\ref{actioneinsteinaether2}), restricted by (\ref{uorthogonal}), and the action (\ref{horavaactiongeneral}) can be seen as follows \cite{Jacobson:2010mx}. The object
\begin{equation}
 P_\alpha{}^\beta = \delta_\alpha{}^\beta + u_\alpha u^\beta 
\end{equation}
is a spatial projector, whereas $P_{\alpha\beta}$ is the induced metric on the spatial hypersurfaces. The extrinsic curvature and the acceleration vector are defined, respectively, by
\begin{equation}
 K_{\mu\nu} = P_\mu{}^\alpha \nabla_\alpha u_\nu \,,
 \hspace{2em}
 a_\mu = u^\alpha \nabla_\alpha u_\mu \,.
\end{equation}
Since $u_\mu$ is hypersurface orthogonal $K_{\mu\nu}$ is a symmetric tensor. $K_{\mu\nu}$ and $a_\mu$ are spatial objects, $K_{\mu\nu} u^\nu = a_\mu u^\mu = 0$. We may decompose the covariant derivative of $u_\mu$ in terms of these objects,
\begin{equation}
 \nabla_\mu u_\nu = K_{\mu\nu} - u_\mu a_\nu \,.
\end{equation}
Now, Since the $T$ field equation need not be imposed explicitly and this a theory with general covariance, we can take $T$ as the time coordinate, $T = t$. By doing so we break the symmetry of general diffeomorphisms over the spacetime. In addition, we can write the spacetime metric in the ADM variables $N$, $N_i$ and $g_{ij}$. With these settings we have that the aether part of the Lagrangian in (\ref{actioneinsteinaether2}) takes the form
\begin{equation}
 M^{\alpha\beta\gamma\delta} \nabla_\alpha u_\gamma \nabla_\beta u_\delta =
 ( c_1 + c_3 ) K_{ij} K^{ij} + c_2 K^2 - ( c_1 - c_4 ) a_i a^i \,.
\end{equation}
In addition, $\sqrt{-g} = \sqrt{^{(3)}\!g} N$ and the decomposition of $R$ adds $K_{ij} K^{ij} - K^2 +\,^{(3)}\!R$ to the Lagrangian. By putting all this in the action (\ref{actioneinsteinaether2}), we have that the $z=1$ Ho\v{r}ava action (\ref{horavaactiongeneral}) is reproduced from it if the coupling constants of both theories are identified according to
\begin{equation}
 \beta = \frac{1}{1 - c_1 - c_3} \,,
 \hspace{2em}
 \kappa_{\mbox{\tiny H}} = \beta \kappa_{\mbox{\tiny EA}} \,,
 \hspace{2em}
 \lambda = \beta ( 1 + c_2 ) \,,
 \hspace{2em}
 \alpha = \beta ( c_1 - c_4 ) \,.
\label{relationsconstants}
\end{equation}
Therefore, the $z=1$ Ho\v{r}ava action (\ref{horavaactiongeneral}) is a gauge-fixed version of the hypersurface-orthogonal Einstein-aether action given in (\ref{actioneinsteinaether2}) and (\ref{uorthogonal}).

\section{Linearized theory}

\subsection{Gauge invariants with the $T$ field}
Now we focus on the linearized generally-covariant theory. Minkowski spacetime, whose metric we denote by $\eta_{\alpha\beta}$, is a solution of the theory in absence of matter sources and with the condition $T = t$, which yields a zero aether energy-momentum tensor. We introduce the perturbative variables by expanding around this solution in the way
\begin{equation}
g_{\mu\nu}(t,\vec{x}) = \eta_{\mu\nu} + \epsilon\, h_{\mu\nu}(t,\vec{x}) \,,
\hspace{2em}
T(t,\vec{x}) = t + \epsilon\, \tau(t,\vec{x}) \,.
\label{covariantperturbations}
\end{equation}
The dependence of  $h_{\mu\nu}$ and $\tau$ on the spacetime coordinates is arbitrary, except for the asymptotic conditions $h_{\mu\nu} , \tau \rightarrow 0$ as $r \rightarrow \infty$.

We investigate the possible gauge invariants of the linearized theory that can be formed with the metric $h_{\mu\nu}$ and the $\tau$ field. Under an arbitrary diffeomorphism over the spacetime, given by
\begin{equation}
\delta x^\mu = \zeta^\mu(x^\alpha) \,,
\end{equation}
the exact spacetime metric $g_{\mu\nu}$ and the exact scalar field $T$ transform as
\begin{eqnarray}
&&  \delta g_{\mu\nu} = 
- \zeta^\alpha \partial_\alpha g_{\mu\nu}
- 2 g_{\alpha(\mu} \partial_{\nu)} \zeta^\alpha \,,
\\
&&  \delta T = -\zeta^\alpha \partial_\alpha T \,.
\end{eqnarray}
On the perturbative variables these transformations take the form
\begin{eqnarray}
&& \delta h_{\mu\nu} = 
- 2 \partial_{(\mu} \zeta_{\nu)} \,,
\label{lineartransfcovh}
\\
&& \delta \tau = \zeta_0 \,,
\label{lineartransfcovtau}
\end{eqnarray}
where we have used the background metric to lower the index, $\zeta_\mu = \eta_{\mu\nu} \zeta^{\nu}$. For the compatibility with the asymptotic conditions on the field variables we require that $\zeta^\mu \rightarrow 0$ as $r \rightarrow 0$.

Now it is convenient to introduce the transverse and longitudinal decompositions for $\zeta_i$, $h_{0i}$ and $h_{ij}$. They are given by
\begin{eqnarray}
&& \zeta_i = \xi_i + \partial_i \chi \,,
\label{decomposezetai}
\\
&& h_{0i} = m_i + \partial_i b \,,
\label{decomposeh0i}
\\
&& h_{ij} =
h_{ij}^{TT} + \frac{1}{2} \left( \delta_{ij} - \partial_{ij} \partial^{-2} \right) h^T + \partial_{(i} h_{j)}^L + \partial_{ij} \partial^{-2} h^L \,.
\label{decomposehij}
\end{eqnarray}
The symbol $\partial_{ij\cdots k}$ stands for $\partial_i \partial_j \cdots \partial_k$, $\partial^2$ is the flat Euclidean Laplacian, $\partial^2 \equiv \partial_{kk}$, and $\partial^{-2}$ is its inverse, $\partial^{-2} \equiv (\partial^2)^{-1}$.  The restrictions on the variables are $\partial_i \xi_i = \partial_i m_i = \partial_i h^L_i = \partial_i h_{ij}^{TT} = h^{TT}_{ii} = 0$. For the uniqueness of the decompositions and the compatibility with the asymptotic behavior of the original field variables, we asume the asymptotic conditions
\begin{equation}
 \xi_i , \chi, m_i, b, h_{ij}^{TT}, h^T, h_i^L, h^L \rightarrow 0 
 \hspace{2em} \mbox{as} \hspace{2em}
 r \rightarrow \infty \,.
\end{equation}
By substituting (\ref{decomposezetai} - \ref{decomposehij}) in the transformation (\ref{lineartransfcovh}), we obtain that it becomes
\begin{eqnarray}
&& \delta h_{00} = - 2 \dot{\zeta}_0 \,,
\label{transh00}
\\
&& \delta b = - \dot{\chi} - \zeta_0 \,,
\\
&& \delta m_i = - \dot{\xi}_i \,,
\\
&& \delta h^L = - 2 \partial^2 \chi \,,
\\
&& \delta h^L_i = - 2 \xi_i \,,
\\
&& \delta h^T = 0 \,,
\\
&& \delta h^{TT}_{ij} = 0 \,.
\label{transhijTT}
\end{eqnarray}
From these transformations we extract that $h^T$ and $h^{TT}_{ij}$ are gauge invariants. By combining with (\ref{lineartransfcovtau}), we may define three variables that are also gauge invariants, namely
\begin{eqnarray}
&& p \equiv h_{00} + 2 \dot{\tau} \,,
\label{p}
\\
&& q \equiv \dot{h}^L - 2 \partial^2 b - 2 \partial^2 \tau \,,
\\
&& v_i \equiv \dot{h}^L_i - 2 m_i \,.
\label{nui}
\end{eqnarray}
This approach is rather different to the standard approach of linearized GR (see, for example, \cite{Flanagan:2005yc}), since the scalar $\tau$ is involved in the definition of the gauge invariants $p$ and $q$. The gauge invariant of linearized GR  that depends only on the metric components is
\begin{equation}
\Phi \equiv \partial^2 h_{00} + \ddot{h}^L - 2 \partial^2 \dot{b} \,.
\label{dependentinvariant}
\end{equation}
Here $\Phi$ is not an independent quantity since it can be obtained from a combination of $p$ and $q$, $\Phi = \partial^2 p + \dot{q}$.
Therefore, in the linearized theory that depends on $h_{\mu\nu}$ and $\tau$ and that is generally covariant, the independent gauge invariants are $h^T$, $h^{TT}_{ij}$, $p$, $q$ and $v_i$. In Appendix A we show that an analogous construction of gauge invariants can be done for the case of the theory formulated with the FDiff-gauge symmetry.

\subsection{Linearized Einstein equations}
Here we study the linearized field equations. We consider the presence of matter sources, hence there is an active energy-momentum tensor $T_{\mu\nu}^{\mbox{\tiny matter}}$ for the matter. We define the energy-momentum tensors $T_{\mu\nu}^{\mbox{\tiny aether}}$ and $T^{\mbox{\tiny matter}}_{\mu\nu}$ in such a way that the Einstein equations take the form 
\begin{equation}
G_{\mu\nu} = 
T_{\mu\nu}^{\mbox{\tiny aether}} 
+ \kappa_{\mbox{\tiny EA}} T^{\mbox{\tiny matter}}_{\mu\nu} \,,
\label{einsteingeneral}
\end{equation}
with the usual expression $G_{\mu\nu} \equiv R_{\mu\nu} - \frac{1}{2} g_{\mu\nu} R$.
We decompose $T^{\mbox{\tiny matter}}_{\mu\nu}$ in the way
\begin{eqnarray}
&& T_{00}^{\mbox{\tiny matter}} = \rho \,,
\label{decomposet00}
\\
&& T_{0i}^{\mbox{\tiny matter}} = S_i + \partial_i S \,,
\\
&& T_{ij}^{\mbox{\tiny matter}} = 
\sigma_{ij}^{TT} 
+ \frac{1}{2} \left( \delta_{ij} - \partial_{ij} \partial^{-2} \right) 
  \sigma^T
+ \partial_{(i} \sigma_{j)}^L
+ \partial_{ij} \partial^{-2} \sigma^L \,,
\label{decomposetij}
\end{eqnarray}
where the variables are restricted by $ \partial_i S_i = \partial_i \sigma_i^L = \partial_i \sigma_{ij}^{TT} = \sigma_{ii}^{TT} = 0$. 

The linearized Einstein equations can be completely expressed in terms of the gauge invariants defined in the previous section. Indeed, after the decompositions (\ref{decomposeh0i} - \ref{decomposehij}) and (\ref{decomposet00} - \ref{decomposetij}) are done and the gauge invariants (\ref{p} - \ref{nui}) are introduced, the linearized Einstein equations take the form
\begin{eqnarray}
 && \beta \partial^2 h^T - \alpha \partial^2 p = 
 - 2 \kappa_{\mbox{\tiny H}} \rho \,,
 \label{eq00}
 \\
 && \lambda \dot{h}^T - ( 1 - \lambda ) q = - 2 \kappa_{\mbox{\tiny H}} S \,,
 \label{long0i}
 \\
 && \partial^2 v_i = 4 \kappa_{\mbox{\tiny H}} S_i \,,
 \label{trans0i}
 \\
 && \lambda \ddot{h}^T - ( 1 - \lambda ) \dot{q} = 
 - 2 \kappa_{\mbox{\tiny H}} \sigma^L \,,
 \label{longlongij}
 \\
 && \dot{v}_i = 2 \kappa_{\mbox{\tiny H}} \sigma_i^L \,,
 \label{longij}
 \\
 && ( 1 - 3 \lambda ) ( \ddot{h}^T + \dot{q} ) + \beta \partial^2 h^T
 - 2 \beta \partial^2 p = 
 2 \kappa_{\mbox{\tiny H}} ( \sigma^T + \sigma^L ) \,,
 \label{traceij}
 \\
 && \ddot{h}^{TT}_{ij} - \beta \partial^2 h^{TT}_{ij} =
 2 \kappa_{\mbox{\tiny H}} \sigma_{ij}^{TT} \,.
 \label{transtraceij}
\end{eqnarray}
Equation (\ref{eq00}) is the $00$ component of the Einstein equations, Eqs.~(\ref{long0i}) and (\ref{trans0i}) come from the $0i$ components and the last four equations constitute the $ij$ components. We have used relations (\ref{relationsconstants}) to change the constants $\kappa_{\mbox{\tiny EA}}$ and $c_{1,2,3,4}$ of the generally-covariant formulation by the constants $\kappa_{\mbox{\tiny H}}$, $\lambda$, $\beta$ and $\alpha$  of the FDiff-covariant formulation since the latter are the ones that the linearized theory naturally adopts. We stress that no gauge-fixing condition has been imposed to obtain these equations. All the variables of the left-hand sides belong to the set of gauge invariants of the linearized theory. In Appendix A we show that if one uses the original FDiff-invariant formulation of the $z=1$ Ho\v{r}ava theory, then the linearized field equations can also be written purely in terms of the corresponding FDiff-gauge invariants. 

Evidently, Eqs.~(\ref{long0i}), (\ref{trans0i}) (\ref{longlongij}) and (\ref{longij}) imply the following conditions on the source,
\begin{eqnarray}
  && \sigma^L = \dot{S} \,,
  \label{preservation1}
  \\
  && \partial^2 \sigma_i^L = 2 \dot{S}_i  \,.
  \label{preservation2}
\end{eqnarray}
Consequently, we drop Eqs.~(\ref{longlongij}) and (\ref{longij}) out from the list of independent Einstein equations and impose Eqs.~(\ref{preservation1} - \ref{preservation2}) as complementary conditions that must be satisfied by the matter source (equations (\ref{preservation1} - \ref{preservation2}) are independent of $h_{\mu\nu}$ and $\tau$).

Let us analyze the system of equations (\ref{eq00} - \ref{transtraceij}) as a set of equations for the gauge invariants with the matter source given and, momentarily, without imposing any restriction on the coupling constants. Equations (\ref{eq00}), (\ref{long0i}) and (\ref{trans0i}) do not depend on the second time derivative of any variable. Hence, they are constraints on the initial data. Specifically, Eq.~(\ref{eq00}) corresponds to the Hamiltonian constraint whereas Eqs.~(\ref{long0i}) and (\ref{trans0i}) constitute the momentum constraint (see the canonical formulation for the theory at the vaccum in \cite{varios:hamiltoniannokc} for the case out of the KCP and in \cite{Bellorin:2013zbp} for the case at the KCP). On the other hand, Eqs.~(\ref{traceij}) and (\ref{transtraceij}) do depend on the second time derivate, so they are the ones that govern the propagation of the dynamical modes. In this covariant formalism we have the ten components of the metric field and the scalar field $\tau$, which sum up eleven field variables. Four of these must be fixed by a coordinate-system choice and the remaining seven are subject to the system (\ref{eq00} - \ref{transtraceij}), which has just seven independent equations (we recall that Eqs.~(\ref{longlongij}) and (\ref{longij}) have been dropped out and that Eqs.~(\ref{trans0i}) and (\ref{transtraceij}) yield four independent equations). Therefore, the system is closed for the field variables with the matter source given. The seven gauge-independent variables are subject to four constraints, Eqs.~(\ref{eq00} - \ref{trans0i}), hence in this general theory there remain three propagating physical degrees of freedom, whose evolution is governed by Eqs.~(\ref{traceij}) and (\ref{transtraceij}). One can identify two of these propagating modes as the two tensorial modes of GR. The remaining mode is an extra scalar mode.

Now we move to the case of our interest. It is evident that Eq.~(\ref{traceij}) changes its character of evolution equation if we set the coupling constant $\lambda$ to
\begin{equation}
 \lambda = 1/3 \,.
 \label{lambdaonethird}
\end{equation}
In this case Eq.~(\ref{traceij}) lacks its dependence on the second time derivative $\ddot{h}^T$, hence it becomes an additional constraint. Notice that the change of evolution equations by constraints is not a smooth one. The kinetic-conformal theory is an independent theory on its own. In terms of the Einstein-aether constants this condition is $c_1 + c_3 + 3 c_2 = -2$. Let us write here the resulting set of independent field equations under the condition (\ref{lambdaonethird}),
\begin{eqnarray}
 && \beta \partial^2 h^T - \alpha \partial^2 p = 
 - 2 \kappa_{\mbox{\tiny H}} \rho \,,
\label{eq00onethird}
\\
 && \dot{h}^T - 2 q = - 6 \kappa_{\mbox{\tiny H}} S \,,
\label{long0ionethird}
\\
 && \partial^2 v_i = 4 \kappa_{\mbox{\tiny H}} S_i \,,
\label{trans0ionethird}
\\
 && \beta \partial^2 h^T - 2 \beta \partial^2 p = 
 2 \kappa_{\mbox{\tiny H}} ( \sigma^T + \dot{S} ) \,,
\label{traceijonethird}
\\
 && \ddot{h}^{TT}_{ij} - \beta \partial^2 h^{TT}_{ij} =
 2 \kappa_{\mbox{\tiny H}} \sigma_{ij}^{TT} \,.
\label{transtraceijonethird}
\end{eqnarray}
The four first equations are constraints that fix the gauge invariants $h^T$, $p$, $q$ and $v_i$, whereas the last one is the evolution equation for the tranverse-traceless tensorial mode $h_{ij}^{TT}$. Therefore, under the condition (\ref{lambdaonethird}), which defines the kinetic-conformal point, the extra mode is annihilated, in agreement with the vacuum theory \cite{Bellorin:2013zbp,Bellorin:2016wsl}, and the propagating physical degrees of freedom are the same of GR, which are described by $h_{ij}^{TT}$.

By imposing some bounds on the coupling constants $\alpha$ and $\beta$, we can ensure that the constraints form a closed system of partial differential equations and that the evolution equation for $h_{ij}^{TT}$ is the sourced wave equation. The conditions on the coupling constants are
\begin{equation}
 \beta > 0 \,,
 \hspace{2em}
 \alpha \neq 2 \beta \,.
\end{equation}
Under these conditions the formal solutions of the constraints (\ref{eq00onethird} - \ref{traceijonethird}) are
\begin{eqnarray}
 h^T &=& 
 - \frac{2 k_{\mbox{\tiny H}}}{ \beta ( 2 \beta - \alpha )}
   \partial^{-2} \left[
       2 \beta \rho + \alpha ( \sigma^T + \dot{S} ) \right] \,,
 \label{solht}
 \\
 p &=& 
 - \frac{2 k_{\mbox{\tiny H}}}{ 2 \beta - \alpha }
   \partial^{-2} \left[ \rho + \sigma^T + \dot{S} \right] \,,
 \\
 q &=& 
 - \frac{ k_{\mbox{\tiny H}} }{ \beta ( 2 \beta - \alpha )}
  \partial^{-2} \left[ 
    2 \beta \dot{\rho} + \alpha ( \dot{\sigma}^T + \ddot{S} ) \right]
 + 3 k_{\mbox{\tiny H}} S \,,
 \label{solq}
 \\
 v_i &=&
 4 k_{\mbox{\tiny H}} \partial^{-2} \left[ S_i \right] \,.
 \label{solnui}
\end{eqnarray}

Equations (\ref{transtraceijonethird}) and (\ref{solht} - \ref{solnui}) tell us that $h^T$, $p$, $q$ and $v_i$ are variables bounded to the sources whereas $h_{ij}^{TT}$ is the radiative variable. Therefore, as in GR, $h_{ij}^{TT}$ acquires a pure physical meanning as the only radiative field.

The speed of the waves $h_{ij}^{TT}$ is $\sqrt{\beta}$. This agrees with Ref.~\cite{Jacobson:2004ts} in what concerns the $h_{ij}^{TT}$ mode. However, as we have already mentioned, in the unrestricted Einstein-aether theory studied in Ref.~\cite{Jacobson:2004ts} there are three additional modes propagating themselves with wave equations and with different speeds. In Ref.~\cite{Foster:2006az} the same linearized equations were studied with matter sources. We remark again that the total equivalence between the hypersurface orthogonal Einstein-aether theory and the nonprojectable $z=1$ Ho\v{r}ava theory holds only if the condition of hypersurface orthogonality is imposed at the level of the action. Therefore, the linearized $z=1$ Ho\v{r}ava equations at the kinetic-conformal point, whose covariant version is (\ref{eq00onethird} -  \ref{transtraceijonethird}), are not obtained by direct substitution of the hypersurface orthogonality condition on the equations analyzed in Refs.~\cite{Jacobson:2004ts,Foster:2006az}.

\subsection{The quadrupole formula}
In the standard perturbative scheme (post-Minkowskian approach), the exact solution is increasely approximated by the perturbative solution if the nolinear terms of the field equations are considered as sources for the fields at the order of interest. For example, the perturbative wave equation (\ref{transtraceijonethird}) at higher order in perturbations can be casted as
\begin{equation}
 \ddot{h}^{TT}_{ij} - \beta \partial^2 h^{TT}_{ij} =
 \left( 2 \kappa_{\mbox{\tiny H}} T_{ij}^{\mbox{\tiny matter}} + t_{ij} \right)^{TT} \,,
\end{equation}
where $h_{ij}^{TT}$ in the left-hand side is evaluated at the order of interest. $t_{ij}$ represents the nonlinear terms coming from the Einstein tensor and the aether energy-momentum tensor. The solutions for all the variables at lower orders must be substituted in $t_{ij}$ and $T_{ij}^{\mbox{\tiny matter}}$ such that the desired order in perturbations is reached in all terms of this equation.

In this iterative scheme the linearized field equations determine the leading contribution. Here we extract information from the solution of the linearized wave equation relevant for the physics at large distances from the source. The solution of Eq.~(\ref{transtraceijonethird}) with no incoming radiation is \begin{equation}
 h_{ij}^{TT} =
 \frac{ \kappa_{\mbox{\tiny H}} }{ 2 \pi \beta }
 \int d^3x' 
 \frac{ \sigma^{TT}_{ij}( t - |\vec{x} - \vec{x}\,'|/\sqrt{\beta} \,, \vec{x}\,' ) }
                 { | \vec{x} - \vec{x}\,' | } \,.
\label{exactsolution}
\end{equation}
Following the outline of Ref.~\cite{Will:1993ns}, our next steps consist of approximating this expression according to weakness criteria: the observer is far from the source, the self-gravity of the source is negligible and the motion of the source is sufficiently slow. Then, by defining $\hat{n} = \vec{x}/r$ with $r = |\vec{x}|$, solution (\ref{exactsolution}) can be expanded in the form
\begin{equation}
 h_{ij}^{TT} =
 \frac{ \kappa_{\mbox{\tiny H}} }{ 2 \pi \beta r }
 \sum\limits_{m=0}^{\infty} \frac{1}{ m! (\sqrt{\beta})^m }
 \frac{\partial^m}{\partial t^m} \int d^3x'
 \left( \hat{n} \cdot \vec{x}\,' \right)^m
 \sigma^{TT}_{ij}( t - r/\sqrt{\beta} \,, \vec{x}\,' ) \,.
\end{equation}
The leading mode of this expansion is
\begin{equation}
 h_{ij}^{TT} =
 \frac{ \kappa_{\mbox{\tiny H}} }{ 2 \pi \beta r }
 \int d^3x' \sigma^{TT}_{ij}( t - r/\sqrt{\beta} \,, \vec{x}\,' ) \,.
\end{equation}
Let us massage this expression following a procedure similar of the one of Ref.~\cite{Foster:2006az}. One can handle first the integration (and the expansion already done) in terms of the full energy-momentum tensor and then perform the projection to the transverse-traceless sector. At order $\mathcal{O}(1/r)$ the projection to the transverse sector is equivalent to the algebraic projection to the plane orthogonal to $\hat{n}$. The projector to this plane is $\theta_{ij} \equiv \delta_{ij} - n^i n^j$, and the operator that projects tensors and substracts the trace is $P^{TT}_{ijkl} \equiv \theta_{ik} \theta_{jl} - \frac{1}{2} \theta_{ij} \theta_{kl}$. Then we have
\begin{equation}
 h_{ij}^{TT} =
 \frac{ \kappa_{\mbox{\tiny H}} }{ 2 \pi \beta r} P^{TT}_{ijkl}
 \int d^3x' 
 T^{\mbox{\tiny matter}}_{kl}( t - r/\sqrt{\beta} \,, \vec{x}\,' )  \,.
\end{equation}
Similarly to the criterium of order used Ref.~\cite{Foster:2006az}, we require that the equations of motion of the matter source, approximated to the order of interest, imply the Newtonian law of mass conservation, namely\footnote{For example, for a perfect fluid, at the nonrelativistic limit it holds $\partial^2 S = - \partial_i ( \rho v^i)$, where $\rho$ is the mass density and $v^i$ the velocity. Then Eq.~(\ref{massconservation}) becomes the nonrelativistic law of mass conservation of the fluid, $\dot{\rho} + \partial_i ( \rho v^i ) = 0$, used in Ref.~\cite{Foster:2006az}.}
\begin{equation}
  \dot{\rho} = \partial^2 S  \,.
  \label{massconservation}
\end{equation}
This and Eqs.~(\ref{preservation1} - \ref{preservation2}) are equivalent to the preservation of the matter energy-momentum tensor, $\partial_\mu T^{\mu\nu}_{\mbox{\tiny matter}} = 0$. Therefore, the requisite (\ref{massconservation}) is equivalent to demand that, at the linear order in perturbations and at the nonrelativistic limit, the equations of motion of the matter sources imply the conservation of its energy-momentum tensor (actually, the spatial components $\partial_\mu T_{\mbox{\tiny matter}}^{\mu i} = 0$, which are given by Eqs.~(\ref{preservation1}) and (\ref{preservation2}), are already implied by the Einstein equations). Following the standard approach of GR, from the conservation of the matter energy-momentum tensor at the order considered one obtains the relation
\begin{equation}
 \int d^3x' T^{\mbox{\tiny matter}}_{ij} =
 \frac{1}{2} \int d^3x' x'^i x'^j \ddot{T}_{00}^{\mbox{\tiny matter}} =
 \frac{1}{2} \int d^3x' x'^i x'^j \ddot{\rho} \equiv \frac{1}{2}\ddot{I}_{ij} \,.
\end{equation}
Therefore, the leading contribution for the generation of gravitational waves has the same structure of Einstein's quadrupole formula that arises in GR,
\begin{equation}
 h_{ij}^{TT} = \frac{ \kappa_{\mbox{\tiny H}}}{ 4 \pi \beta r } 
 P^{TT}_{ijkl} \frac{d^2 I_{kl}( t - r/\sqrt{\beta} ) }{dt^2}  \,.
 \label{quadrupole}
\end{equation}
The only difference this formula has with respect to Einstein's quadrupole formula is the presence of the coupling constants $\kappa_{\mbox{\tiny H}}$ and $\beta$ of the Ho\v{r}ava or Einstein-aether theory. If these constants are adjusted to the GR values $\kappa_{\mbox{\tiny H}} = 8 \pi G_{\mbox{\tiny N}}$ and $\beta = 1$, then (\ref{quadrupole}) becomes identical to Einstein's quadrupole formula.


\section{On the observational bounds}
In this section we discuss some observational bounds on the theory at the kinetic-conformal point. Some of the formulas we use can be directly deduced from the analysis previously done in the literature of the nonprojectable Horava theory with general $\lambda$. However, there are features that do not follow as a particular case of the general theory, essentially due to the discontinuity in the number of degrees of freedom.

We comment that in the (projectable and nonprojectable) Ho\v{r}ava theory with $\lambda \neq 1/3$ there are theoretical restrictions on the coupling constants that are necessary for the stability of the extra mode. Although these conditions are widely used, they do not apply in the kinetic-conformal formulation due to the obvious reason that there is no extra mode. For example, there is a restriction on $\lambda$ given by \cite{Blas:2009qj}
\begin{equation}
 \frac{ 3 \lambda - 1 }{ \lambda - 1 } > 0,
\end{equation}
necessary to avoid that the extra mode becomes a ghost at the level of the linearized theory. The interpretation in the kinetic-conformal formulation, $\lambda = 1/3$, is simply that there is no such bound. We highlight this point since the bounds coming from the physics of the extra mode in the case $\lambda \neq 1/3$ are frequently combined with the observational bounds when the phenomenology of the theory in under scrutiny (see, for example, \cite{Barausse:2011pu,Yagi:2013ava}), hence extrapoling directly the conclusions from the $\lambda \neq 1/3$ case can be misleading in the kinetic-conformal case. 

We start with the observational bounds of the weak regime englobed in the parametrized-post-Newtonian (PPN) parameters of the solar-system tests. The PPN parameters of the kinetic-conformal theory can be obtained from the general nonprojectable Ho\v{r}ava theory with arbitrary $\lambda$ since the discrepancy in the propagating degrees of freedom does not affect the PPN potentials. The PPN parameters for the nonprojectable Ho\v{r}ava theory were computed in Ref.~\cite{Blas:2011zd} (see also \cite{Blas:2010hb}), using the covariant formulation of the second-order in derivatives action, i.~e., the hypersurface-orthogonal Einstein-aether theory. The PPN parameters are obtained for a weak, non-relativistic source, and the procedure is similar to the one of the Einstein-aether theory \cite{Foster:2005dk}. It results that the hypersurface orthogonal Einstein-aether theory (as well as the unrestricted Einstein-aether theory) reproduces the same values of the PPN constants of GR, except for the parameters $\alpha_1$ and $\alpha_2$, whose nonzero values signal violations of the Lorentz symmetry. The expressions obtained in \cite{Blas:2011zd} for these constants, written with our conventions for the coupling constants and for general $\lambda$, are
\begin{eqnarray}
 \alpha_1 &=& 8 ( \beta - 1 ) - 4 \alpha  \,,
 \label{alpha1}
 \\
 \alpha_2 &=& 
 \left[ \frac{ \beta ( 1 - \lambda ) + ( \beta - 1 ) ( 1 - 3 \lambda )
 	      - \alpha ( 1 - 2 \lambda ) }
 	      { 4 ( 2 \beta - \alpha ) ( 1 - \lambda ) } \right] \alpha_1 \,.
\end{eqnarray}
Notice that $\alpha_1$ is independent of $\lambda$. For $\lambda = 1/3$, the parameter $\alpha_2$ becomes
\begin{equation}
 \alpha_2 = \frac{1}{8} \alpha_1 \,.
 \label{alpha2}
\end{equation}
The current observational bounds on these parameters, which are dimensionless, are $|\alpha_1| < 10^{-4}$ and $|\alpha_2| < 10^{-7}$ \cite{Will:2014kxa}. In the kinetic-conformal theory the relation (\ref{alpha2}) demands that the strong bound, which is the one on $\alpha_2$, must be satisfied by both parameters. Taking this into account, we use relation (\ref{alpha1}) to solve one coupling constant in terms of the other one, 
\begin{equation}
 \alpha = 2 ( \beta - 1 ) + \delta \,,
 \label{restriction1}
\end{equation}
where $\delta$ represents the narrow observational window for the $\alpha_2$ parameter, i.~e.~$|\delta| < 10^{-7}$.
With (\ref{restriction1}) the kinetic-conformal Ho\v{r}ava theory satisfies all the conditions derived from the PPN analysis of the solar-system tests.

Nonrelativistic gravitational theories (coupled to relativistic particles) produce Cherenkov radiation if the velocities of the gravitational modes are lower than the speed of the relativistic particles \cite{Moore:2001bv}. The implications of this have been studied for the Einstein-aether theory in Refs.~\cite{Elliott:2005va,Barausse:2011pu}, obtaining very stringent lower bounds on the coupling constants. In the  Einstein-aether theory the lower bounds affect several coupling constants since this theory has several propagating modes, each one with a different dependence of its velocity on the coupling constants. In the kinetic-conformal Ho\v{r}ava theory the lower bounds coming from the Cherenkov radiation are very simple to implement since the propagating modes are the same of GR. We have seen that, when the theory is truncated to its second-order effective action, the squared velocity of the transverse-traceless tensorial modes of the linearized theory is $\beta$, hence in this case the bounds affect only to this constant. Therefore, the Cherenkov radiation puts the bound
\begin{equation}
 \beta \geq 1 \,.
\end{equation}
We comment that at higher energies the higher order operators could be relevant for this analysis, hence other coupling constants can enter in the game.

Now we want to make some considerations about the kinetic-conformal theory at cosmological scales, since there is an important restriction on this theory at this scale. The restriction concerns to the Lagrangian formulation of the theory, but in the Hamiltonian formalism it is still an open question \cite{Bellorin:2017gzj}. In the field equations derived from the Lagrangian, if the ansatz for a full homogeneous and isotropic metric is imposed, together with a homogeneous and isotropic perfect fluid, then it turns out that the only possibility left by the field equations is that the density and pressure vanish. With full homogeneous and isotropic we mean that these conditions are imposed in all the components of the spacetime metric. This restriction can be deduced from the analysis of Ref.~\cite{Blas:2009qj}, where the extension of the nonprojectable theory and its first cosmological application were presented (the Einstein-aether theory exhibits an analogous behavior \cite{Mattingly:2001yd,Carroll:2004ai}). There it was found that the effective cosmological gravitational constant arising in the Friedmann equations differs from the Newtonian (local) gravitational constant by a scale factor that depends on $\lambda$. At the kinetic-conformal point, $\lambda = 1/3$, this scale factor diverges. What really this divergence means is the vanishing of the density and the pressure at the kinetic-conformal point, as we have commented. We point out two issues concerning this restriction. The first one is that in the kinetic-conformal theory the equations of motion derived from the Hamiltonian admits more solutions than the Lagrangian field equations \cite{Bellorin:2017gzj}. This is essentially due to the role played by the Lagrange multipliers once all the constraints have been added to the Hamiltonian. On certain configurations admissible for the Lagrange multipliers, the Legendre transformation cannot be inverted, hence the Hamiltonian lost the equivalence with the Lagrangian. The second issue is that the Ho\v{r}ava theory is originally formulated as a theory with the symmetry of the FDiff, it is not a generally covariant theory. Indeed, the equivalence with the hypersurface orthogonal Einstein-aether theory holds only at the level of the action of second-order in derivatives and on the Lagrangian formulations. Under the reduced FDiff symmetry, the conditions of homogeneity and isotropy can be restricted to the spatial metric, whereas the laspe function and the shift vector can have a more general dependence on the time and the space. These two observations have been discussed in Ref.~\cite{Bellorin:2017gzj}. Moreover, the analysis of the cosmological-scale configurations in the kinetic-conformal Ho\v{r}ava gravity requires a previous and deep analysis on the structure of the second-class constraints of the theory. Perhaps a reformulation of the theory only in terms of first-class constraints is convenient. Under such scenario, some of the original second-class constraints could be circumvented by recurring to different ``gauge fixing" conditions. Thus, the configurations of cosmological scale would require a better gauge fixing than, for example, the $\pi = 0$ condition that arises originally as a second-class constraint. Therefore, we consider that the kinetic-conformal point, $\lambda = 1/3$, is not necessarily ruled out by the restriction about homogenous and isotropic configurations mentioned above.

Finally, we contrast with the bounds coming from binary pulsars, which are strong sources, that were obtained in Refs.~\cite{Yagi:2013qpa,Yagi:2013ava} on two Lorentz-violating theories: the Einstein-aether theory and the hypersurface-orthogonal Einstein aether theory without the kinetic-conformal condition (called the krhonometric theory in those references). The authors of \cite{Yagi:2013qpa,Yagi:2013ava} arrive at very stringent constraints on the space of coupling constants of these theories after contrasting with the observations on several binary pulsars. For the khronometric theory in particular, the region of the space of parameters on which the theory can reproduce the decay rate of orbital period within the observational error excludes the kinetic-conformal point $\lambda = 1/3$. This is because they combine the analysis on the binary pulsar with the bounds resulting from the stability of the extra mode, the Cherenkov radiation and the cosmological-scale effect of rescaling the gravitational constant. As we have discussed, the stability of the extra mode does not apply in the kinetic-conformal theory. The Cherenkov radiation only constraint the constant $\beta$, one can put $\beta \gtrsim 1$ consistently on the kinetic-conformal theory. The rescaling of the gravitational constant at cosmological scales leads to the apparent restriction at $\lambda = 1/3$ that we discussed above. Upon the arguments we have given we consider that it does not rule out unavoidably the kinetic-conformal formulation of the theory.

Indeed, it is interesting to extrapolate the formulas of the orbital evolution of the binary pulsars obtained in \cite{Yagi:2013qpa,Yagi:2013ava} to the kinetic-conformal case. This orbital evolution is related to the effective multipoles of the theory, and we have shown in the previous sections that in the kinetic-conformal theory the dominant mode in the far zone of a weak source is the same of GR. We reproduce here the rate of change of the orbital period obtained in \cite{Yagi:2013ava}, which uses previous results of \cite{Foster:2006az,Foster:2007gr,Blas:2011zd},
\begin{equation}
\begin{array}{rcl}
 {\displaystyle \frac{\dot{P}_b}{P_b}} &=&
 {\displaystyle 	
 -\frac{ 3 a G_{\mbox{\tiny \AE}} }{ \mathcal{G} \mu m } 
 \left< \left\{ 
 \frac{\mathcal{A}_1}{5} \dddot{Q}_{ij} \dddot{Q}_{ij} 
 + \frac{\mathcal{A}_2}{5} \dddot{\mathcal{Q}}_{ij} \dddot{Q}_{ij}
 + \frac{\mathcal{A}_3}{5} \dddot{\mathcal{Q}}_{ij} \dddot{\mathcal{Q}}_{ij}
 \right. \right. }
 \\[2ex] && {\displaystyle
	\left.\left.
 + \mathcal{B}_1 \dddot{I} \dddot{I} 
 + \mathcal{B}_2 \dddot{\mathcal{I}} \dddot{I}
 + \mathcal{B}_3 \dddot{\mathcal{I}} \dddot{\mathcal{I}} 
 + \mathcal{C} \dot{\Sigma}_i \dot{\Sigma}_i
 \right\} \right> } \,.
\end{array}
\label{orbitalperiodgeneral}
\end{equation}
$\mathcal{G}$ is the effective gravitational constant in the binary system, $G_{\mbox{\tiny\AE}}$ is the gravitational constant of the theory (related to $\kappa_{\mbox{\tiny EA}}$ and $\kappa_{\mbox{\tiny H}}$), $a$ is the semi-major axis, $m \equiv m_1 + m_2$, and $\mu \equiv m_1 m_2 / m$. The quadrupole moment $Q_{ij}$ is the trace-free part of the system's mass quadrupole moment $I_{ij}$:
\begin{equation}
 I_{ij} = \sum\limits_A m_A x_A^i x_A^j \,.
\end{equation}
$\mathcal{Q}_{ij}$ is the trace-free part of the rescaled mass quadrupole moment $\mathcal{I}_{ij}$:
\begin{equation}
 \mathcal{I}_{ij} = \sum\limits_A s_A m_A x_A^i x_A^j \,,
\end{equation}
where $s_A$ are constants associated to the sensitivities of the system, which are parameters encoding the departing of bodies' wordlines from the relativistic trayectories. The dipolar moment $\Sigma^i$ is
\begin{equation}
 \Sigma^i = - \sum\limits_A s_A m_A v_A^i \,.
\end{equation}
We have written only the dominant modes for the multipoles. Further details can be found in \cite{Yagi:2013ava,Foster:2006az,Foster:2007gr,Blas:2011zd}. The coefficients arising in (\ref{orbitalperiodgeneral}), using our conventions for the coupling constants, are given by
\begin{equation}
\begin{array}{ll}
 {\displaystyle
 \mathcal{A}_1 \equiv \frac{1}{c_t} 
 + \frac{3 \alpha ( \mathcal{Z} - 1 )^2}{2 ( 2 \beta - \alpha ) c_s} \,, } &
 {\displaystyle
 \mathcal{A}_2 \equiv 
 - \frac{ 2 \beta ( \mathcal{Z} - 1 ) }{ (2 \beta - \alpha ) c_s^3 } \,, }
 \\ {\displaystyle
 \mathcal{A}_3 \equiv 
 \frac{2 \beta^2 }{ 3 \alpha ( 2 \beta - \alpha ) c_s^5 } \,, }
 & {\displaystyle
 \mathcal{B}_1 \equiv 
 \frac{ \alpha \mathcal{Z}^2 }{ 4 ( 2 \beta - \alpha ) c_s } \,, }
 \\ {\displaystyle
 \mathcal{B}_2 \equiv 
 - \frac{ \beta \mathcal{Z} }{ 3 ( 2 \beta - \alpha ) c_s^2 } \,, }
 & {\displaystyle 
 \mathcal{B}_3 \equiv 
 \frac{\beta^2}{ 9 \alpha ( 2 \beta - \alpha ) c_s^5 } \,, }
 \\ {\displaystyle
 \mathcal{C} = 
 \frac{4 \beta^2 }{ 3 \alpha ( 2 \beta - \alpha ) c_s^3 } \,, }
\end{array}
\label{coefficients}
\end{equation}
where 
\begin{equation}
 \mathcal{Z} \equiv 
 \frac{ \alpha_1 - 2 \alpha_2 }{ 3 ( 2 \beta - \alpha - 2 )}
 = \frac{ 2 \alpha_2 }{ 2 \beta - \alpha - 2 } \,,
\end{equation}
and
\begin{equation}
 c_t^2 = \beta \,, \quad
 c_s^2 = 
 \frac{ \beta ( 2 \beta - \alpha ) \lambda }{ \alpha ( 3 \lambda - 1 ) } \,,
 \label{velocity}
\end{equation}
are the velocities of the tensorial modes and the extra scalar mode respectively. Notice that $\mathcal{A}_1$ is the only coefficient with information about the propagation of the tensorial modes. In (\ref{orbitalperiodgeneral}) the angled-brackets stand for an average over several wavelengths. 

Now, it is easy to see how all the contributions of the extra mode disappear and only the quadrupole contribution remains as the dominant mode. By substituting $\lambda = 1/3$ in the velocity $c_s$ given in (\ref{velocity}), we get that $c_s$ diverges, which is an informal way to express that the extra mode gets frozen. Then all coefficients in (\ref{coefficients}) vanish except for $\mathcal{A}_1$, which becomes $\mathcal{A}_1 = 1/\sqrt{\beta}$. Equation (\ref{orbitalperiodgeneral}) becomes
\begin{equation}
 \frac{\dot{P}_b}{P_b} =
 -\frac{ 3 a G_{\mbox{\tiny \AE}} }{ 5 \sqrt{\beta} \mathcal{G} \mu m } 
 \left< \dddot{Q}_{ij} \dddot{Q}_{ij} \right>
\,.
\end{equation}
This results expresses that in the kinetic-conformal theory the quadrupole mode is the dominant radiative contribution to the rate of the orbital period decay, as in GR. Again, we may put $\beta = 1$ consistently in the kinetic-conformal theory.


\section*{Conclusions}
Our central results are that, first, in the Lorentz-violating, power-counting renormalizable (once the higher order operators are considered) and unitary theory coupled to matter sources that we have considered, the only radiative degrees of freedom are the same transverse-traceless tensorial modes of GR, and that, second, at the leading order the Einstein quadrupole formula is reproduced (if two coupling constant are adjusted to their GR values). We have worked in the gauge-invariant formalism of the linearized theory, hence our results are not gauge-fixing artifacts. We have clearly identified the set of the gauge-invariant variables that are nonradiative and the ones that are radiative. The nonradiative variables are linked to the sources by Poissonian equations.

It is interesting that the kinetic-conformal Ho\v{r}ava theory is able to reproduce the same leading mode of GR in what concerns the production and propagation of gravitational waves, but with a better quantum behavior. In Ref.~\cite{Bellorin:2016wsl} we showed the power-counting renormalizability of this theory without matter sources by analyzing the superficial degree of divergence of general one-particle irreducible diagrams, as well as its unitarity. There remains to prove its complete renormalizability.

Of course, a study of the next post-Newtonian orders is needed to make an exhaustive comparison between the gravitational waves produced and propagated in the kinetic-conformal Ho\v{r}ava theory and the detected signals. Our study is the first step towards this goal since it is restricted to the leading order that can be extracted from a weak source. At higher orders, the other field variables besides $h_{ij}^{TT}$ are also relevant for the wave production and propagation due to the nonlinearities of the theory.

We have also considered some observational bounds. The PPN analysis fixes one of the coupling constants of the action of second-order in derivatives, such that all solar-system tests are satisfied by the theory after this restriction is imposed. Cherenkov radiation puts a lower bound on other coupling constant. The $\lambda = 1/3$ value that defines the kinetic-conformal condition is not discarded by the bounds coming from binary pulsars showed in \cite{Yagi:2013qpa,Yagi:2013ava}, since these bounds were combined with theoretical restrictions and cosmological considerations that do not necessarily apply to our case, as we have argued. Indeed, we have shown, based on the formulas of \cite{Yagi:2013ava}, how the rate of decay of orbital period is very close to its corresponding expression in GR at the dominant level. In particular, there is no dipolar contribution to this decay, unlike the theories considered in \cite{Yagi:2013qpa,Yagi:2013ava}.


\section*{Acknowledgments}
A. R. is partially supported by grant Fondecyt No.~1161192, Chile. J. B. is partially supported by the Programa MECE Educaci\'on Superior of Ministerio de Educaci\'on, Chile.

\appendix
\section{Waves in the FDiff formulation}
\subsection{The FDiff gauge invariants}
\label{sec:fdiffformalism}
In this appendix we study the formulation of linearized gauge invariants when the gauge symmetry is given by the diffeomorphisms that preserve a given foliation, the FDiff.

Since there is a subtlety when implementing the FDiff transformations together with a prescribed asymptotic-flatness condition on the field variables, let us start by presenting the FDiff without asymptotic conditions. The diffeomorphisms that preserve a given foliation act on the coordinates $(t,\vec{x})$ in the following way
\begin{equation}
\delta t = f(t) \,,
\hspace{2em}
\delta x^i = \zeta^i(t,\vec{x}) \,.
\end{equation}
The corresponding transformation on the ADM variables is
\begin{equation}
\begin{array}{l}
\delta N = \zeta^k \partial_k N + f \dot{N} + \dot{f} N \,,
\\[1ex]
\delta N_i = \zeta^k \partial_k N_i + N_k \partial_i \zeta^k 
+ f \dot{N}_i + \dot{f} N_i + \dot{\zeta}^j g_{ij} \,,
\\[1ex]
\delta g_{ij} = \zeta^k \partial_k g_{ij} + 2 g_{k(i} \partial_{j)} \zeta^k 
+ f \dot{g}_{ij}  \,.
\end{array}                
\label{fdiff}
\end{equation}

Minkowski spacetime is a vacuum solution of the theory without cosmological constant. We introduce the perturbative variables in the following way
\begin{equation}
N = 1 + \epsilon n \,,
\hspace{2em}
N_i = \epsilon n_i \,,
\hspace{2em}
g_{ij} = \delta_{ij} + \epsilon h_{ij} \,.
\end{equation}

Since the transformation parameters $f$ and $\zeta^i$ are of linear order in perturbations, it is convenient to redefine them in the way $f(t) \rightarrow \epsilon f(t)$ and $\zeta^i(t,\vec{x})\rightarrow \epsilon \zeta^i(t,\vec{x})$, such that the new variables $f$ and $\zeta^i$ do not depend on the scale of perturbations $\epsilon$. The FDiff transformations of the perturbative variables are
\begin{eqnarray}
&& \delta n = \dot{f} \,,
\label{transformationperturbative1}
\\
&& \delta n_i = \dot{\zeta}_i \,,
\\
&& \delta h_{ij} = 2 \partial_{(i} \zeta_{j)} \,. 
\label{transformationperturbative3}
\end{eqnarray}
Now, we impose the asymptotic conditions needed for asymptotic flatness. We require that $n,n_i,h_{ij} \rightarrow 0$ as $r \rightarrow 0$, and require that the parameters $f(t)$ and $\zeta^i(t,\vec{x})$ be compatible with these conditions. However, as we anticipated, there is a subtlety here since $f(t)$ is a function only of time. If we require that $f(t)$ goes to zero as $r \rightarrow \infty$, then necessarily $f(t) = 0$. The consequence for the linearized theory is that the perturbative variable $n$ is actually a gauge invariant under linearized FDiff, $\delta n = 0$. Besides this, we require $\zeta_i \rightarrow 0$ as $r \rightarrow \infty$.

More information about the FDiff transformations (\ref{transformationperturbative1} - \ref{transformationperturbative3}) is extracted when the vectors $\zeta_i$ and $n_i$ and the tensor $h_{ij}$ are decomposed in transverse and longitudinal parts. We introduce the standard decompositions
\begin{eqnarray}
&& \zeta_i = \xi_i + \partial_i \chi \,,
\label{decompositionzeta}
\\
&& n_i = m_i + \partial_i b \,,
\\
&& h_{ij} =
h_{ij}^{TT} + \frac{1}{2} \left( \delta_{ij} - \partial_{ij} \partial^{-2} \right) h^T + \partial_{(i} h_{j)}^L + \partial_{ij} \partial^{-2} h^L \,.
\label{decompositionhij}
\end{eqnarray}
The vectors and tensors of the above decomposition are restricted by $\partial_i \xi_i = \partial_i m_i = \partial_i h_i^L = \partial_i h_{ij}^{TT} = h_{kk}^{TT} = 0$.

By performing the transverse and longitudinal decomposition on the transformations (\ref{transformationperturbative1} - \ref{transformationperturbative3}), we obtain the decomposed FDiff gauge transformations, namely
\begin{eqnarray}
&& \delta n = 0 \,,
\\
&& \delta b = \dot{\chi} \,,
\\
&& \delta m_i = \dot{\xi}_i \,,
\\
&& \delta h^L = 2 \partial^2 \chi \,,
\\
&& \delta h^L_i = 2 \xi_i \,,
\\
&& \delta h^T = 0 \,,
\\
&& \delta h^{TT}_{ij} = 0 \,.
\end{eqnarray}
Actually, these transformations can be deduced directly from (\ref{transh00} - \ref{transhijTT}) by setting $\zeta_0 = 0$, which we recall is required on the FDiff transformations by the asymptotic conditions. From these transformations we automatically extract that $n$, $h^T$ and $h_{ij}^{TT}$ are FDiff-gauge invariant. The combinations
\begin{eqnarray}
&& Q \equiv \dot{h}^L - 2 \partial^2 b \,,
\\
&& v_i \equiv \dot{h}_i^L - 2 m_i 
\end{eqnarray}
are also gauge invariants. 

It is illustrating to contrast with the gauge invariants of linearized pure GR (without the $T$ or aether field). In linearized GR the gauge symmetry is bigger, so we expect having less gauge invariants on the side of GR when the same field variables are used (ADM variables, in this case). Indeed, if the gauge transformation is extended to a general spacetime diffeomorfism, as in (\ref{transh00} - \ref{transhijTT}), then the variables $h^T$, $h_{ij}^{TT}$ and  $v_i$ are still gauge invariants. $n$ and $Q$ are not separately invariant, but the combination of them, 
\begin{equation}
 \Phi = - 2 \partial^2 n + \dot{Q} \,,
\end{equation}
is. This is the invariant of GR defined in (\ref{dependentinvariant}). There are seven functional degrees of freedom among the FDiff-gauge invariants $h^T$, $h_{ij}^{TT}$, $v_i$, $n$ and $Q$, whereas there are six gauge invariants on the side of pure GR since the enhacement of the gauge symmetry leads to the lacking of one of them, only a combination of $Q$ and $n$ survives.

\subsection{The linearized field equations}
\label{sec:fdiffdynamics}
In this appendix we study the dynamics of the linearized theory in the FDiff-covariant formalism. For simplicity we consider here the pure gravity theory, without coupling to matter sources. The action of second order in derivatives is (\ref{horavaactiongeneral}),
\begin{equation}
S = \frac{1}{2\kappa_{\mbox{\tiny H}}} \int dt d^3x \sqrt{g} N \left (
G^{ijkl} K_{ij} K_{kl} + \beta R + \alpha a_i a^i \right) \,.
\end{equation}
The equations of motion, obtained by taking variations with resptect to $g_{ij}$, $N$ and $N_i$, are, respectively,
\begin{eqnarray}
\frac{1}{\sqrt{g}} \frac{\partial}{\partial t} 
( \sqrt{g} G^{ijkl} K_{kl} )
+ 2 G^{klm(i|} \nabla_k ( K_{lm} N^{|j)} )
- G^{ijkl} \nabla_m (  K_{kl} N^m )
& &\nonumber \\
+ 2 N ( K^{ik} K_k{}^j - \lambda K K^{ij} )
- \frac{1}{2} N g^{ij} G^{klmn} K_{kl} K_{mn}  
& & \nonumber \\ 
+ \beta N (R^{ij} - \frac{1}{2} g^{ij} R )
- \beta ( \nabla^i \nabla^j N - g^{ij} \nabla^2 N )
&& \nonumber \\
+ \alpha N^{-1} ( \nabla^i N \nabla^j N 
- \frac{1}{2} g^{ij} \nabla_k N \nabla^k N )
& = & 0 \,,
\label{eomgij}
\\
G^{ijkl} K_{ij} K_{kl} - \beta R 
+ 2 \alpha N^{-2} ( N \nabla^2 N - \frac{1}{2} \nabla_i N \nabla^i N ) 
& = & 0 \,,
\label{eomn}
\\
G^{ijkl} \nabla_j K_{kl} &=& 0 \,.
\label{eomni}
\end{eqnarray}

Equations (\ref{eomn}) and (\ref{eomni}) are constraints on the initial data since they do not contain second-order time derivatives of the field variables. These equation are the analogous of the Hamiltonian and momentum constraints of GR. The additional constraint arising at the kinetic-conformal point $\lambda =1/3$ can be extracted from Eq.~(\ref{eomgij}). Indeed, with $\lambda = 1/3$ the hypermatrix $G^{ijkl}$ becomes degenerated,
\begin{equation}
g_{ij} G^{ijkl} = 0 \,,
\label{singularmatrix}
\end{equation}
$g_{ij}$ being its null eigenvector. In Eq.~(\ref{eomgij}) the only term that has a second-order time derivative is the first one, specifically when $\frac{\partial}{\partial t}$ acts on $K_{kl}$. It is clear that if we take the trace of this equation with $g_{ij}$, considering $\lambda = 1/3$, this second-order time derivative disappears due to (\ref{singularmatrix}). Thus, at $\lambda =1/3$ the trace of Eq.~(\ref{eomgij}) is another constraint of the theory. It can be written in the form
\begin{equation}
N G^{ijkl} K_{ij} K_{kl} 
- G^{ijkl} \nabla_i K_{jk} N_l
+ (\frac{1}{2} + \lambda) \left( K_{ij} \nabla^i N^j 
- K \nabla_k N^k \right)
- (\beta - \frac{\alpha}{2} ) \nabla^2 N = 0 \,.
\label{constraintc}
\end{equation}
We have used the constraint (\ref{eomn}) to bring the trace to this form. Therefore, at $\lambda = 1/3$ the constraints of the theory, in the Lagrangian formalism, are the Eqs.~(\ref{eomn}), (\ref{eomni}) and (\ref{constraintc}).

Now we study the field equations (\ref{eomgij}), (\ref{eomn}), (\ref{eomni}) and (\ref{constraintc}) perturbatively. It turns out that they can be completely expressed and solved in terms of the FDiff-gauge invariants introduced in the previous appendix. At linear order in perturbations, constraints (\ref{eomn}) and (\ref{constraintc}) yield, respectively,
\begin{eqnarray}
\beta \partial^2 h^T + 2 \alpha \partial^2 n &=& 0 \,,
\\
( 2\beta - \alpha ) \partial^2 n &=& 0 \,.
\end{eqnarray}
By assuming $\alpha \neq 2\beta$, $\beta \neq 0$ and the asymptotic behavior $h^T = \mathcal{O}(1/r)$ and $n = \mathcal{O}(1/r)$ at $r\rightarrow \infty$, we have that these equations imply $h^T = n = 0$ at linear order in perturbations. The perturbative version of the constraint (\ref{eomni}), after substituting $h^T = 0$ into it, takes the form
\begin{equation}
 \partial^2 v_i + 2 ( 1 - \lambda ) \partial_i Q = 0 \,.
 \label{perturbativemomentumcons}
\end{equation}
We recall that we are considering $\lambda = 1/3$. The spatial divergence of this equation yields the condition 
\begin{equation}
 \partial^2 Q = 0 \,.
\label{eqc}
\end{equation}
Assuming the asymptotic behavior $Q = \mathcal{O}(1/r)$ at $r\rightarrow \infty$, this equations has $Q = 0$ as its only solution. Putting this back in Eq.~(\ref{perturbativemomentumcons}) we obtain $v_i = 0$ if $v_i = \mathcal{O}(1/r)$ at spatial infinity.

Finally, we study the perturbative version of the field equation (\ref{eomgij}), droping out all the variables that we already known are zero. It yields the wave equation for $h_{ij}^{TT}$,
\begin{equation}
\ddot{h}^{TT}_{ij} - \beta \partial^2 h^{TT}_{ij} = 0 \,.
\end{equation}

Summarizing, the linearized FDiff formulation also admits a representation in terms of FDiff gauge invariants. The field equations are completely analogous to the generally-covariant formulation (in the vacuum, in this case). The fundamental result is that $h_{ij}^{TT}$ is the only propagating mode and that it is radiative (in the sense of the linearized theory). We remark that the vanishing of the gauge invariants $h^T$, $n$, $Q$ and $v_i$ holds only in the vacuum theory. If matter sources were present these variables were nonzero and their expressions were bounded to the sources in the sense we presented in the generally-covariant formulation.


\end{document}